\documentclass[aps,prl,
       reprint,
       superscriptaddress,
       showpacs,
   ]{revtex4-1}
\usepackage{amsfonts}
\usepackage{amssymb}
\usepackage{amsmath}
\usepackage{color}
\usepackage{graphicx}
\usepackage{subfigure}
\usepackage[pdftex,colorlinks=true,linkcolor=red,citecolor=blue]{hyperref}
\usepackage{color}
\usepackage{epsfig}
\usepackage{graphicx}
\usepackage{time}
\newcommand{\rd}{\mathrm{d}}               
\newcommand{\rD}{\mathrm{D}}               
\newcommand{\defby}{\equiv}                
\newcommand{\calN}{\mathcal{N}}            
\newcommand{\calL}{\mathcal{L}}            
\newcommand{\Expect}[1]
   {\ensuremath{\langle \, #1 \,  \rangle}}
\newcommand{\Comm}[2]
   {\ensuremath{[ \, #1, #2 \, ]}}
\newcommand{\AntiComm}[2]
   {\ensuremath{\{ \, #1, #2 \, \}}}
\newcommand{\Tr}[1]{\mathrm{Tr} [ \, #1 \, ]}      
\newcommand{\Ln}[1]{\ln [ \, #1 \, ]}              
\begin{document}
%
%
%

%
%

\preprint{LA-UR-10-05099; \today}

\title[BEC]
   {Non-perturbative predictions for cold atom Bose gases with tunable interactions}

\author{Fred Cooper}
\affiliation{
   Los Alamos National Laboratory,
   Los Alamos, NM 87545}
\affiliation{Santa Fe Institute,
   Santa Fe, NM 87501}

\author{Chih-Chun Chien}
\affiliation{
   Los Alamos National Laboratory,
   Los Alamos, NM 87545}

\author{Bogdan Mihaila}
\affiliation{
   Los Alamos National Laboratory,
   Los Alamos, NM 87545}

\author{John F. Dawson}
\affiliation{Department of Physics,
   University of New Hampshire,
   Durham, NH 03824}

\author{Eddy Timmermans}
\affiliation{
   Los Alamos National Laboratory,
   Los Alamos, NM 87545}


\begin{abstract}
We derive a theoretical description for dilute Bose gases as a loop expansion in terms of composite-field propagators by rewriting the Lagrangian in terms of auxiliary fields related to the normal and anomalous densities.  We demonstrate that already in leading order this non-perturbative approach  describes a large interval of coupling-constant values, satisfies Goldstone's theorem, yields a Bose-Einstein transition that is second-order, and is consistent with the critical temperature predicted in the weak-coupling limit by the next-to-leading order large-N expansion. 
\end{abstract}

\maketitle

%
%

Nearly a century after the first observation of the lambda transition in liquid helium\cite{r:Kamerling:1911fk}, a quantitative, first-principles description of strongly-correlated bosons remains a challenge. After the transition was recognized as the onset of superfluidity\cite{r:Kapitza:1938kx,*r:Allen:1938vn}, the connection with Bose-Einstein condensation (BEC) was proposed\cite{r:London:1938ys,*r:London:1938zr}, but it was Bogoliubov's work\cite{r:Bogoliubov:1947ys} pointing out that the dispersion of the elementary BEC excitations satisfy the Landau criterion for superfluidity\cite{r:Landau:1941ly} that motivated weakly-interacting BEC studies to investigate superfluid properties.  In weakly-interacting systems, the many-body properties do not depend on the shape of the interaction potential, but only 
on 
the $s$-wave scattering length, $a_{0}$, and the boson fluid acts as point-like interacting particles\cite{r:Lee:1957ve}.

Unlike liquid helium, cold atoms remain point-like even when the scattering length is tuned near a Feshbach resonance.  Then, strongly-correlated cold atom bosons offer the exciting prospect of studying point-like strongly interacting bosons, possibly in the universal regime where the scattering length greatly exceeds the inter-particle distance and the latter becomes the only relevant length scale\cite{r:Shin:2007oq}.  This hope appeared thwarted when it was shown that the three-body loss rate in cold atom traps scales as $a_0^{4}$ near a Feshbach resonance\cite{r:Fedichev:1996hc,*r:Esry:1991ij}. In accordance, the universal regime was reached only in ultra-cold fermionic gases\cite{r:Ho:2004kl,*r:Blume:2007tg}, where the three-body loss is reduced by virtue of the Pauli exclusion principle.  However, the recent observation that three-body losses are strongly suppressed in optical lattices when the average number of bosons per site is two or less\cite{r:Daley:2009bs}, rekindles the prospect of studying medium and strongly-correlated cold atom bosons. 
Novel cold-atom trap technologies that produce stable, flat potentials bound by a sharp edge\cite{r:Henderson:2006fv,*r:Henderson:2009dz}, suggest the study of finite-temperature properties such as the BEC transition temperature $T_c$ and the superfluid to normal fluid ratio and depletion, at fixed density, $\rho$.

At finite temperature, the description of BEC's remains a challenge even in the weakly-interacting regime.  Standard approximations such as the Hartree-Fock-Bogoliubov and the Popov schemes, generally fall within the Hohenberg and Martin classification\cite{r:Hohenberg:1965fu} of conserving and gapless approximations, which implies that they either violate Goldstone's theorem or general conservation laws\cite{r:Griffin:1996kl}.  These approximations generally predict the BEC transition to be a first-order transition, whereas we expect the transition to be second order\cite{r:Andersen:2004uq}.

In this paper, we present a new theoretical framework that describes a large interval of $\rho^{1/3}a_0$-values, satisfies Goldstone's theorem and yields a Bose-Einstein transition that is second-order, while also predicting reasonable values for the depletion.  
\textcolor{black}{Furthermore, this framework can predict \emph{all} experimentally relevant quantities within the same calculation, determining fully consistently quantities such as  $T_{c}$, the collective mode frequencies\cite{ref:eddyN1} and the compressibility (which characterizes the density profile in a shallow trap\cite{ref:eddyN2}).}
In contrast with other resummation schemes, such as the large-$N$ expansion\cite{r:Baym:2000fk} or functional renormalization techiques\cite{r:Floerchinger:2008kx}, here we treat the normal and anomalous densities on equal footing.  
%

\textcolor{black}{
In our approach, we generate a one-parameter family of equivalent Lagrangians. 
We choose this parameter to reproduce the one-loop result at mean-field level in the weakly-interacting limit. 
Thus, we identify the optimal auxiliary-field Lagrangian for the purpose of a systematic non-perturbative expansion.
Then, the critical temperature variation in leading order is the same as the one found in the next-to-leading order large-$N$ expansion.
}
%
%

In dilute bosonic gas systems, the classical action 
is given by
$
   S[\, \phi,\phi^{\ast} \, ]
   =
   \int \! \rd x \> 
   \calL[ \, \phi,\phi^{\ast} \, ] 
$, 
with $\rd x \defby \rd t \, \rd^3 x$ and the Lagrangian density
\begin{gather}
   \calL[ \, \phi,\phi^{\ast} \, ]
   =
   \frac{i \hbar}{2} \, 
   [ \, 
      \phi^{\ast}(x) \, ( \partial_t \, \phi(x) )
      -
      ( \partial_t \, \phi^{\ast}(x) ) \, \phi(x) \,
   ]
   \notag \\
   {}-
   \phi^{\ast}(x) \, 
   \Bigl \{ \,
      -
      \frac{\hbar^2\nabla^2}{2m}
      -
      \mu \,
   \Bigr \} \, 
   \phi(x)
   -
   \frac{\lambda}{2} \, | \, \phi(x) |^4 \>.
   \label{BEC.aux.e:LagI}
\end{gather}
Here, $\mu$ is the chemical potential and the coupling is $\lambda = 4\pi \hbar^2 \, a_0 / m$.
To account for the contributions of the normal and anomalous densities, 
we use the Hubbard-Stratonovitch transformation\cite{r:Hubbard:1959kx,*r:Stratonovich:1958vn}
to introduce the real and complex auxiliary fields (AF),  $\chi(x)$ and $A(x)$. 
We add to Eq.~\eqref{BEC.aux.e:LagI} the AF Lagrangian density\cite{r:Bender:1977bh,r:Coleman:1974ve,*r:Root:1974qf}
\begin{align}
   &\calL_{\text{aux}}[\phi,\phi^{\ast},\chi,A,A^{\ast}]
   =
   \frac{1}{2 \lambda} \,
   \bigl [ \,
      \chi(x) - \lambda \, \cosh\theta \, | \phi(x) |^2 \,
   \bigr ]^2
   \notag \\
   & \qquad{}-
   \frac{1}{2 \lambda} \,
   \bigl | \,
      A(x) 
      - 
      \lambda \, \sinh\theta \, \phi^{2}(x) \,
   \bigr |^2
   \>,
   \label{BEC.aux.e:Laux}
\end{align}
where $\theta$ is the mixing parameter between the normal and anomalous densities, $\chi(x)$ and $A(x)$.  The usual large-N approximation\cite{r:Coleman:1974ve,*r:Root:1974qf} is obtained when $\theta = 0$.  Then, the action becomes
\begin{align}
   &S[\Phi,J] =S[ \phi_a,\chi,A,A^\ast, j_a,s,S]
   \label{BEC.aux.e:actionII} \\
   & \quad 
   =
   - \frac{1}{2} \, 
   \iint \rd x \, \rd x' \,
   \phi_a(x) \, G^{-1}{}^a{}_b[\chi,A](x,x') \, \phi^b(x')
   \notag \\
   & \qquad {}+
   \int \rd x \,
   \bigl \{ \,
      \bigl [ \,
         \chi^2(x) - | A(x) |^2 \,
      \bigr ] / (2\lambda)
      -
      s(x) \chi(x)
      \notag \\
      &
      +
      S^{\ast}(x) A(x)
      +
      S(x) A^{\ast}(x)
      +
      j^{\ast}(x) \phi(x)
      +
      j(x) \phi^{\ast}(x) \,
   \bigr \} \>,
   \notag
\end{align}
with
\begin{align}
   &G^{-1}{}^a{}_b[\chi,A]
   = 
   \bigl \{ \,
      G^{-1}_0{}^a{}_b  
      +
      V^{a}{}_{b}[\chi,A](x) \,
   \bigr \}  \delta(x,x') \, \>,
   \notag \\
   &G^{-1}_0{}^a{}_b
   =
   \begin{pmatrix}
      h_0 & 0 \\[3pt]
      0 & h_0^{\ast} 
   \end{pmatrix} \>,
   \quad
   h_0
   =
   -
   \frac{\hbar^2 \nabla^2}{2m}
   -
   i \hbar \frac{\partial}{\partial t}
   -
   \mu \>,   
   \label{BEC.aux.e:G0invdef} \\
   &V^{a}{}_{b}[\chi,A](x)
   =
   \begin{pmatrix}
      \chi(x) \cosh\theta & - A(x) \sinh\theta \\
      - A^{\ast}(x) \sinh\theta & \chi(x) \cosh\theta
   \end{pmatrix} \>.
   \notag
\end{align}
Here, we introduced a two-component notation with $\phi^a(x) = \{ \, \phi(x), \phi^{\ast}(x) \, \}$ for $a = 1,2$.  
$\Phi(x)$ and $J(x)$ signify the five-component fields and currents.  
%
%
The generating functional for connected graphs is
\begin{equation*}\label{BEC.Seff.e:Z}
   Z[J]
   =
   e^{i W[J] / \hbar}
   = 
   \calN
   \int \rD \Phi \>
   e^{ i S[\Phi;J] / \hbar } \>,
\end{equation*}
with $S[\Phi;J]$ given by Eq.~\eqref{BEC.aux.e:actionII}.  
Performing the path integral over the fields $\phi_a$, we obtain the effective action 
\begin{align*}
   &\epsilon \,S_{\text{eff}}[ \chi;J,\epsilon ]
   =
   \frac{1}{2 } \iint \rd x \, \rd x' \,
   j_{a}(x) \, G[\chi]^a{}_b(x,x') \, j^a(x)
   \\
   &{}+
   \int \rd x \,
   \Bigl \{ \,
      \frac{\chi_i(x) \, \chi^{i}(x)}{2\lambda}
      -
      S_{i}(x) \, \chi^{i}(x)
      - \frac{\hbar}{2i}  
      \text{Tr} \, \Ln{ G^{-1} } \, 
   \Bigr \} \>,
\end{align*}
where $
   \chi^{i}(x)
   =
   \bigl \{ 
      \chi(x), A(x)/\sqrt{2}, A^{\ast}(x)/\sqrt{2} 
   \bigr \} 
   ,
   \>
   S^{i}(x)
   =
   \bigl \{ 
      s(x), S(x)/\sqrt{2}, S^{\ast}(x)/\sqrt{2} 
   \bigr \} 
   $.
The  small parameter $\epsilon$ allows us to perform the remaining path integral over $\chi^i$ using the stationary-phase approximation.  
As shown in Ref.\onlinecite{r:Bender:1977bh}, $\epsilon$ counts loops in the AF propagator in analogy with $\hbar$, and provides the loop expansion of the effective action in terms of $\chi$ propagators.  Next, we expand the effective action about the stationary points, $\chi_0^{i}(x)$, defined by $\delta S_{\text{eff}}[ \chi;j ] /  \delta \chi_{i}(x) = 0$.
Hence, we obtain 
\begin{align*}
   \frac{\chi_0(x)}{\lambda}
   &=
   \bigl \{ \,
      | \phi_0(x) |^2
      +
      \frac{\hbar}{2i} \, \Tr{ G(x,x) } \,
   \bigr \} \, \cosh\theta
   + s(x) \>,
  \\
  \frac{A_0(x)}{\lambda}
   &=
   \bigl \{ \,
       \phi^2_0(x) 
      +
      \frac{\hbar}{i} \, G^{2}{}_{1}(x,x) \,
   \bigr \} \, \sinh\theta
   + S(x) \>,
\end{align*}
where we introduced the notations 
\begin{equation*}\label{BEC.Seff.e:phi0def}
   \phi^a_0[\chi_0](x)
   =
   \int \rd x' \, G[\chi_0]^a{}_b(x,x') \, j^b(x') \>.   
\end{equation*}
We emphasize that both $\chi_0$ and $A_0$ include self-consistent fluctuations. 
Expanding the effective action about the stationary point, we write 
\begin{align}
   &S_{\text{eff}}[ \chi;J ]
   =
   S_{\text{eff}}[ \chi_0;J ]
   +
   \frac{1}{2} \iint \rd^4 x \, \rd^4 x' \,
   D_{ij}^{-1}[\chi_0](x,x')
   \notag \\ & \qquad {} \times
   \label{BEC.Seff.e:Seffexpand} 
   [ \chi^i(x) - \chi^i_0(x) ] \,
   [ \chi^j(x') - \chi^j_0(x') ]
   +
   \dotsb
   \>,
\end{align}
where $D_{ij}^{-1}(x,x')$ is given by the second-order derivatives,
\begin{equation*}
   D_{ij}^{-1}[\chi_0](x,x')
   =
   \frac{ \delta^2 \, S_{\text{eff}}[ \chi^a] }
        { \delta \chi^i(x) \, \delta \chi^j(x') } \, \bigg |_{\chi_0} 
   \>,
\end{equation*}
evaluated at the stationary points.  
By keeping the gaussian fluctuations and Legendre transforming, the one-particle irreducible (1-PI) graphs generating functional
\begin{align}
   &\Gamma[\Phi]
   =
   \int \rd x \, j_{\alpha}(x) \, \phi^{\alpha}(x)
   -
   W[J]
   \label{BEC.Seff.e:vertexfctdef} \\
   &
   =
   \frac{1}{2} \iint \rd x \, \rd x' \,
   \phi_a(x) \, G^{-1}[\chi]^{a}{}_{b}(x,x') \, \phi^b(x')
   \notag \\
   &\quad {}-
   \int \rd x \,
   \Bigl \{ \,
      \frac{\chi_{i}(x) \, \chi^{i}(x)}{2\lambda}
      - \frac{\hbar}{2i}  
      \text{Tr} \bigl \{ \,
         \Ln{ G^{-1}[\chi](x,x) } \,
      \bigr \} 
      \notag \\
      &\qquad \qquad \qquad {}-
      \frac{\hbar \, \epsilon}{2i} \,
      \text{Tr} \, \Ln{ D_{ii}^{-1}[\Phi](x,x) } \, 
   \Bigr \}
   +
   \dotsb
   \>,
   \notag    
\end{align}
is the negative of the classical action plus self-consistent one-loop corrections in the $\phi_a$ and $\chi_i$ propagators. 

%
%

To leading order in the AF loop expansion (LOAF), one sets
 $\epsilon = 0$ in the right-hand-side of \eqref{BEC.Seff.e:vertexfctdef}.
The static part of the effective action per unit volume is 
\begin{align}
   V_{\text{eff}}[\Phi]
   &=
   ( 
      \chi \cosh\theta - \mu \, 
   ) \, | \phi |^2
   -
   \frac{1}{2} \,
   (
      A^{\ast} \, \phi^2 
      +
      A \, \phi^{\ast\,2} 
   ) \sinh\theta
   \notag \\
   & \qquad{}-
      \frac{\chi^2 - | A |^2 }{2\lambda}
      + 
      \frac{\hbar}{2i}  
      \text{Tr} \bigl \{ \,
         \Ln{ G^{-1}[\chi ] } \, 
      \bigr \} \>.
   \label{BEC.Seff.e:Veff}    
\end{align}
Translating \eqref{BEC.Seff.e:Veff} to the imaginary time formalism, we find
\begin{equation*}
   \frac{\hbar}{2i}  
   \text{Tr} \, \Ln{ G^{-1}[\chi ] } 
   =
   \int \frac{\rd^3 k}{(2\pi)^3} \,
   \Bigl \{ \,
      \frac{\omega_k}{2}
      +
      \frac{1}{\beta} \,  \Ln{ 1 - e^{-\beta \omega_k} } \,
   \Bigr \} \>,
\end{equation*}
where $\omega_k^2 = ( \epsilon_k + \chi \, \cosh\theta - \mu )^2 -  |A|^2 \sinh^2\theta $ and $\epsilon_k = k^2/(2m)$.
At the minimum, we have
\begin{equation}\label{BEC.Seff.e:brokencase}
   \frac{\delta V_{\text{eff}}[\Phi]}{\delta \phi^{\ast}} \Bigl |_{\phi_0}
   =
   ( \chi \cosh\theta - \mu ) \, \phi_0
   -
   A  \, \sinh\theta \, \phi_0^{\ast}
   =
   0 \>.
\end{equation}
Using the $U(1)$ gauge symmetry, we choose $\phi_0$ to be real. Then, $A$ is real and the dispersion, $\omega_k^2 = \epsilon_k ( \epsilon_k + 2 A \sinh\theta )$, represents the Goldstone theorem.  Next, we set $\sinh\theta = 1$,  such that  $\omega_k$ reduces to the Bogoliubov dispersion, $\omega_k = \sqrt{\epsilon_k ( \epsilon_k + 2 \lambda \, \phi_0^2 ) }$, in the limit of vanishing quantum fluctuations in the anomalous density.  We note that the leading-order (LO) in the large-N expansion corresponds to $\theta = 0$. This leads to the noninteracting (NI) dispersion, $\omega_k = \epsilon_k$, and we conclude that  the large-N expansion is not a suitable starting point, because it is incompatible with  the Bogoliubov spectrum.

Using standard regularization techniques\cite{r:Papenbrock:1999fk}, the renormalized effective potential is written as
\begin{align*}
   &V_{\text{eff}}[\Phi]
   =
   \chi' | \phi |^2
   -
   \frac{1}{2} 
   \bigl ( 
      A^{\ast} \phi^2
      +
      A \phi^{\ast\,2} 
   \bigr ) 
   -
   \frac{ ( \chi' + \mu)^2}{4\lambda}
   +
   \frac{| A |^2 }{2\lambda}
   \\
   & 
   +  \!\!
   \int \!\! 
   \frac{\rd^3 k}{(2\pi)^3} 
   \Bigl [ 
      \frac{1}{2}
      \Bigl (
         \omega_k
         -
         \epsilon_k
         -
         \chi'
         +
         \frac{|A|^2}{2 \epsilon_k} 
      \Bigr )
      \! +
      \frac{1}{\beta}  \ln( 1 - e^{-\beta \omega_k} ) 
   \Bigr ] ,
\end{align*}
where $\chi' \! = \! \sqrt{2} \chi - \mu$ and $\omega_k^2 = \! ( \epsilon_k + \chi' + |A|) ( \epsilon_k + \chi' - |A|)$.  The gap equations, obtained from $\delta V_{\text{eff}}[\Phi] / \delta \chi^{i} = 0$, are
\begin{align}
   \frac{A}{\lambda}
   &=
   \phi^2
   +
   A
   \int \frac{\rd^3 k}{(2\pi)^3} \,
   \Bigl \{ 
      \frac{1 + 2 n(\omega_k)}{2\omega_k}
      -
      \frac{1}{2\epsilon_k}
   \Bigr \} \>,
      \label{BEC.Seff.e:gapeqsA} 
   \\ \notag
   \frac{\chi'+ \mu}{2 \lambda}
   &=
   | \phi |^2
   +
   \int \frac{\rd^3 k}{(2\pi)^3} \,
   \Bigl \{ 
      \frac{\epsilon_k + \chi'}{2\omega_k}
      [ 1 + 2 n(\omega_k) ]
      -
      \frac{1}{2}
   \Bigr \} 
   \>,
\end{align}
where $n(\omega_k)=[ \exp(\omega_k/k_BT) - 1]^{-1} $ is the Bose-Einstein particle distribution.
At the minimum of the effective potential we have, $ (\chi'_0 - A_0 )\, \phi_0 = 0$, see Eq.~\eqref{BEC.Seff.e:brokencase}, 
and we replace $\mu$ by the physical density using
$
   \rho 
   = 
   - \partial V_{\text{eff}}[\Phi_0] / \partial \mu 
   = ( \chi'_0 + \mu ) / ( 2 \lambda ) 
$.
The density is used to rescale Eqs.~\eqref{BEC.Seff.e:gapeqsA}, and the ensuing phase diagram problem depends only on the dimensionless parameter, $\rho^{1/3}a_0$, and the coupling constant becomes $\lambda=8\pi \, \rho^{1/3}a_0$.
In~the broken symmetry phase, we have $\chi'_0 = A_0$ and the dispersion relation, $\omega_k^2 = \epsilon_k ( \epsilon_k + 2 \chi'_0 )$. The condensate density is denoted by $\rho_0 = \phi_0^2$.  
At weak coupling and $T=0$, 
our results coincide with the Bogoliubov  (one-loop) approximation\cite{r:Andersen:2004uq}, 
$
   \mu
   =
   8\pi \rho a_0  
   \bigl [
      1
      +
      (32/3) \sqrt{\rho a_0^{3} / \pi } \,
   \bigr ] 
$.

%
%
\begin{figure}[t!]
   \centering
   \includegraphics[width=0.8\columnwidth]{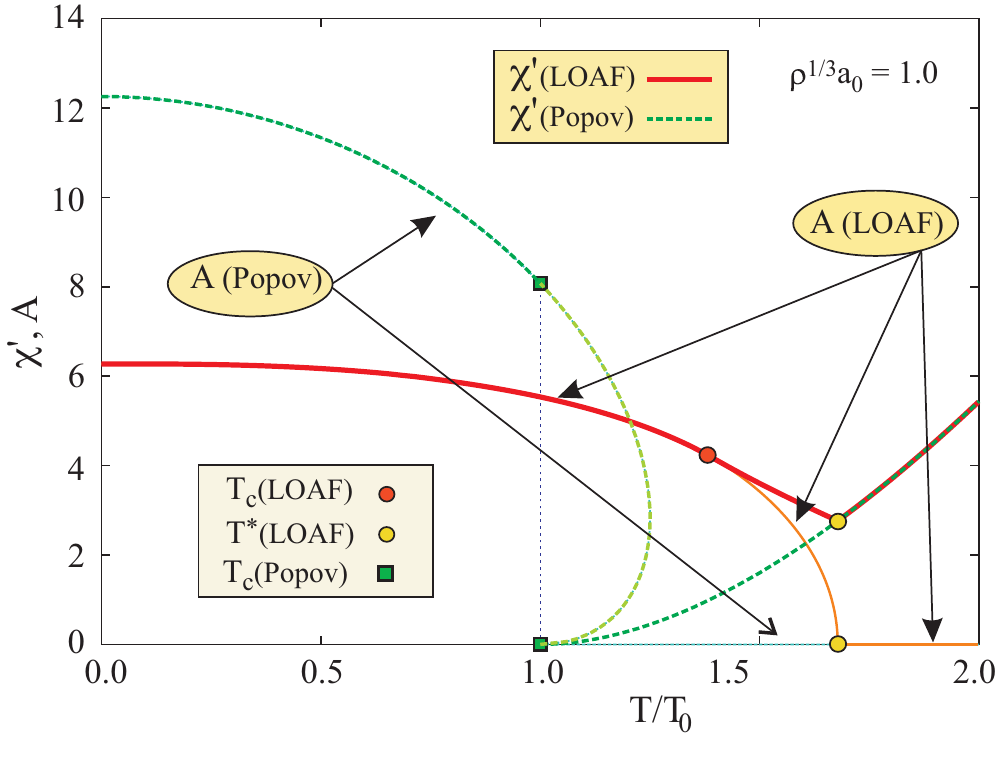}
   \caption{\label{f:densities} (Color online) Normal density, $\chi'$, and anomalous density, $A$, from the LOAF and PA approximations, for $\rho^{1/3}a_0 = 1$. 
   $T_c$ and $T^\star$ indicate vanishing condensate density, $\rho_0$, and anomalous density, $A$, respectively. 
   PA leads to a first-order phase transition, whereas LOAF predicts a second-order phase transition.
   We have that $T_c = T^\star$ in the PA, but not in LOAF. In LOAF $\chi'$ and $A$ are equal until $T_c$.
   }
\end{figure}
%
%
\begin{figure}[h!]
   \centering
   \includegraphics[width=0.8\columnwidth]{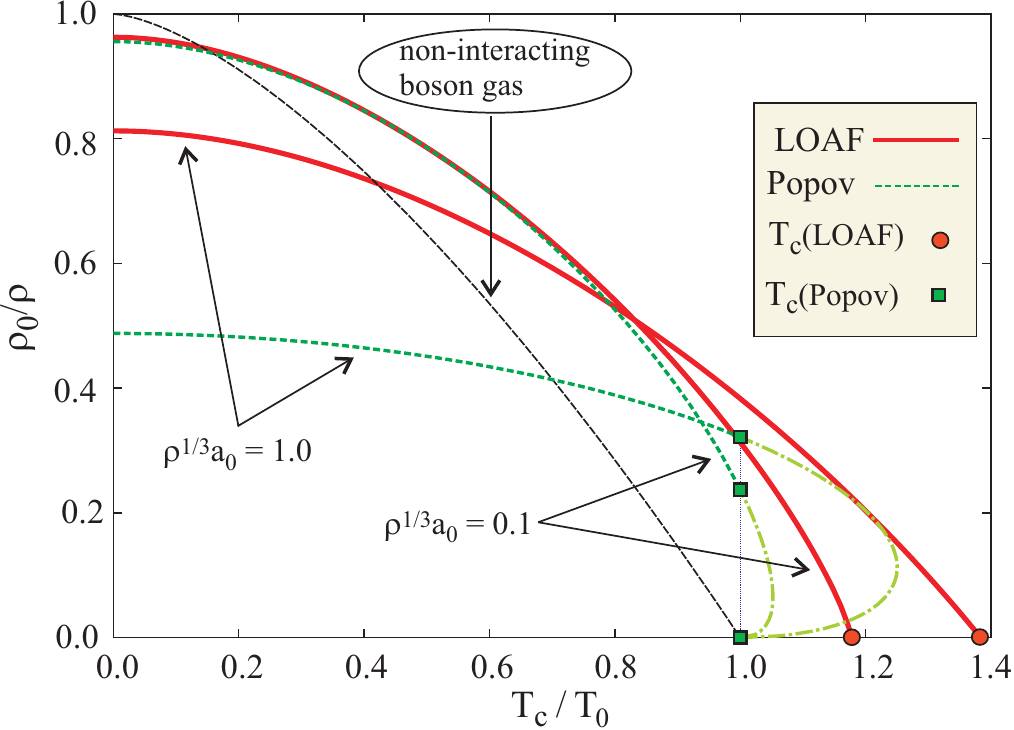}
   \caption{\label{f:condensate}(Color online) Temperature dependence of the condensate fractions from LOAF and PA, 
   compared with the NI result, for $\rho^{1/3}a_0 = 0.1$ and $\rho^{1/3}a_0 = 1$. 
   Because at $T_c$ the PA and NI dispersion relations are the same, PA does not change $T_c$ relative to the NI case.
   LOAF increases $T_c$.}
\end{figure}
%
%
\begin{figure}[h!]
   \centering
   \includegraphics[width=0.8\columnwidth]{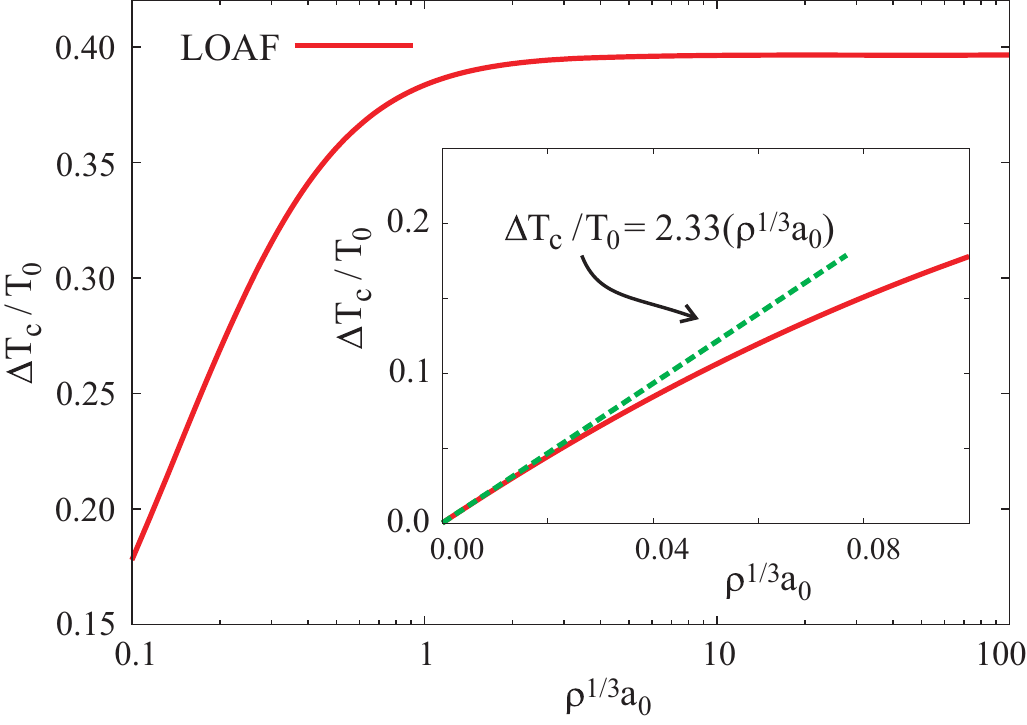}
   \caption{\label{f:deltaT} (Color online) Relative change in $T_c$ with respect to NI, as  predicted by LOAF as a function of $\rho^{1/3}a_0$. 
   The inset shows that in the weak-coupling regime, LOAF produces the same slope as the next-to-leading order large-N expansion\cite{r:Baym:2000fk}.}
\end{figure}
%
%

We compare the LOAF results with the predictions of the Popov bosonic approximation (PA)\cite{r:Popov:1987vn}.  PA is generally recognized as an accurate theoretical description of experimental data in weakly-coupled dilute trapped Bose gases\cite{r:Dalfovo:1999fk}, as long as the densities of the condensed and noncondensed atoms are comparable with each other. Unfortunately, PA produces an artificial first-order phase transition at $T_c$.  
Formally, PA 
is obtained from Eq.~\eqref{BEC.Seff.e:gapeqsA} by setting $A_0 =\chi_0' = \lambda \rho_0$ and neglecting the quantum fluctuations in the anomalous density.  With this substitution, the PA dispersion relation reads $\omega_k^2 = \epsilon_k ( \epsilon_k + 2 \lambda \rho_0 )$. 

In Fig.~\ref{f:densities} we depict the temperature dependence of the normal density $\chi'$, and anomalous density, $A$,  at constant $\rho^{1/3}a_0$, as derived using the LOAF and PA approximations. For illustrative purposes, we set $\rho^{1/3}a_0 = 1$ and the temperature is scaled by its NI critical value, $T_0 = (2\pi \hbar^2 / m) [\rho/\zeta(3/2)]^{2/3}$, where $\zeta(x)$ is the Riemann zeta function. We identify two special temperatures, at $T_c$ where the condensate density vanishes, and at $T^\star$ where the anomalous density, $A$, vanishes. These temperatures are the same in the PA formalism, but they are different in LOAF. 
\textcolor{black}{The existence of a temperature range, $T_c < T < T^\star$, for which the anomalous density, $A$, is nonzero despite a zero condensate fraction, $\phi$, is a fundamental prediction of LOAF. In this temperature range, the dispersion relation is expected to depart from the quadratic form predicted by the Popov approximation for $T > T_c$.}
Above $T_c$ the solution of the PA equations becomes multivalued, indicating that the system undergoes a first-order phase transition at $T_c$. In contrast, LOAF predicts a second-order transition.

The temperature dependence of the condensate fraction, $\rho_0/\rho$, is depicted in Fig.~\ref{f:condensate} for two constant values of the dimensionless parameter $\rho^{1/3}a_0$, together with the NI result, $\rho_0/\rho = 1 - ( T/T_0 )^{3/2}$. Again, we observe that LOAF exhibits the correct second-order BEC phase transition behavior. Moreover, PA does not change $T_c$ relative to the NI case, because in the PA case we have $T_c=T^\star$ and the PA and NI dispersion relations are the same at $T_c$. The LOAF approximation predicts an increase of $T_c$ compared with the NI case.

As illustrated in Fig.~\ref{f:condensate}, the LOAF and PA predictions may differ greatly even for temperatures, $T \ll T_c$. These differences are enhanced by a strengthening of the interaction between particles in the Bose gas (a larger value of $\rho^{1/3}a_0$ indicates stronger coupling). The leading-order AF formalism  produces a more realistic set of observables away from the weak-coupling limit  because of its non-perturbative character. In contrast,  PA is appropriate only in the case of a weakly-interacting gas of bosons. The former is made explicit by studying the LOAF prediction for the relative change in $T_c$ with respect to $T_0$, as a function of $\rho^{1/3}a_0 $. The inset in Fig.~\ref{f:deltaT} demonstrates that in the weak-coupling regime, $\rho^{1/3}a_0 \ll 1$, LOAF produces the same slope of the linear departure derived by Baym \emph{et al.}\cite{r:Baym:2000fk} using the large-N expansion, but at next-to-leading order. 
\textcolor{black}{
The LOAF corrections to the critical temperature are due to the inclusion of self-consistent fluctuations effects in the mean-field $\chi'$ and $A$ densities.
A summary of $\Delta T_c/T_0$ theoretical predictions is found in Ref.\onlinecite{r:Andersen:2004uq}.
For $\rho^{1/3}a_0 \gg 1$, LOAF predicts that $\Delta T_c/T_0  \rightarrow 0.396$ when the system approaches the unitarity limit.
Despite that most current experiments probe only the $\rho^{1/3}a_0 \ll 1$ regime, future experiments\cite{r:Henderson:2006fv,*r:Henderson:2009dz} may access  the medium-to-strongly interacting regime, and verify this non-perturbative prediction.
}

%
%

One can systematically improve upon the LOAF approximation 
by  calculating the 1-PI action order-by-order in~$\epsilon$. The broken $U(1)$ symmetry Ward identities guarantee Goldstone's theorem order by order in $\epsilon$ \cite{r:Bender:1977bh}. 
For time-dependent problems, however, this expansion is secular\cite{r:MCD01}, and a further resummation is required. The latter is performed using the two-particle irreducible (2-PI) formalism\cite{r:Baym62,*r:CJT}. 
A~practical implementation of this approach is the bare-vertex approximation (BVA)\cite{r:BCDM01}. The BVA is an energy-momentum and particle-number conserving truncation of the Schwinger-Dyson infinite hierarchy of equations obtained by ignoring the derivatives of the self-energy, similarly to the Migdal's theorem\cite{r:Migdal:1958uq} approach in condensed matter physics. The BVA proved effective in the case of classical and quantum $\lambda \phi^4$ field theory problems\cite{r:CDM02,*r:CDM02ii,*r:Mihaila:2003ys} and can be applied to the BEC case.

%
%

To summarize, in this paper we introduce a new non-perturbative resummation formulation for the BEC problem. At mean-field level, this approach meets three important criteria for a satisfactory mean-field theory for weakly-interacting bosons\cite{r:Andersen:2004uq}:
i)~the excitation spectrum is gapless (to preserve Goldstone's theorem),
ii)~LOAF reduces to the known results from Bogoliubov theory at $T=0$ and weak coupling, 
and iii)~predicts a second-order BEC phase transition.
The latter suggests that a AF formulation of the Lagrangian for systems of cold fermionic atoms may also impact the study of the BEC to BCS crossover in dilute fermionic atom systems\cite{r:Levin:2010fk}.


Work performed in part under the auspices of the U.S. Department of Energy.  
The authors would like to thank E. Mottola and P.B.~Littlewood for useful discussions.  

%
\bibliography{johns}

\begin{thebibliography}{40}
\bibitem [{ref(){\natexlab{b}}}]{ref:eddyN2}%
  \BibitemOpen
  \href@noop {} {} \bibinfo {note} {In the Thomas-Fermi
  approximation, the local BEC density, $\rho(x)$, in a trapping potential,
  $V_{T}(x)$, follows from the density-dependent chemical potential, $\mu(\rho) = \mu - V_{T}(x)$. 
  Taking the gradient of both sides, we find that
  the local trap force experienced by the bosons $F_{T}(x)=-\nabla V_{T}(x)$
  and the boson density gradient, $\nabla \rho(x)$, are proportional with a
  constant of proportionality equal to $\partial \mu/\partial \rho$, related to
  the compressibility, $\nabla \rho(x)/F_{T}(x)= \delta \mu/\delta \rho$. With
  the sensitive density profile measurement developed for fermion thermometry,
  experimentalists could, in principle, verify the compressibility
  calculation.}\BibitemShut {Stop}%
\end{thebibliography}

\vfill
%
%
\end{document}